# BOUNDARY LAYER FLOW AND HEAT TRANSFER OF CROSS FLUID OVER A STRETCHING SHEET

Masood KHAN, Mehwish MANZUR[1] and Masood ur RAHMAN

Department of Mathematics, Quaid-i-Azam University, Islamabad 44000, Pakistan.

**Abstract:** The current study is a pioneering work in presenting the boundary layer equations for the two-dimensional flow and heat transfer of the Cross fluid over a linearly stretching sheet. The system of partial differential equations is turned down into highly non-linear ordinary differential equations by applying suitable similarity transformations. The stretching sheet solutions are presented via. a numerical technique namely the shooting method and graphs are constructed for the shear-thinning as well as shear-thickening regime. The impact of the emerging parameters namely the power-law index $m$, the local Weissenberg number $We$ and the Prandtl number $\Pr$ on the velocity and temperature fields are investigated through graphs. Numerical values of the local skin friction coefficient and the local Nusselt number are also presented in tabular form. For some limiting cases, comparisons with previously available results in the literature are made and an excellent agreement is achieved.

Keywords: Cross fluid; Stretching sheet; Heat Transfer; Numerical solution.

## 1. Introduction

Nowadays, the generalized Newtonian fluids have gained massive prominence amongst researchers on account of their significant practicality in industrial, chemical and technological mechanisms. In the generalized Newtonian fluids [1-2] the viscosity is shear dependent as a result of which a new constitutive relation is defined by making modifications in the Newton's law of viscosity to account for the change in viscosity with varying shear rate. The most common type of the generalized Newtonian fluids is the power-law fluid [3] which tenders a simplest representation of the shear-thinning/thickening behaviors of several fluids. Despite of the abundant process engineering applications, the main limitation of the power-law fluid is that it cannot describe the fluid behavior for very low and very high shear rates but only for limited range of shear rate called the power-law region. When the deviation from the power-law model is notable only at very low shear rate, Ellis model [4] is utilized. Likewise, Sisko model [5] characterize the flow of fluids in the power-law and very high shear rate region. In order to overcome all the limitations of the above mentioned rheological models, a broader sub-class of the generalized Newtonian fluids namely the Cross model was introduced by Cross [6]. This model is competent of depicting the flow in the power-law region as well as the regions of very low and very high shear rates. Unlike the power-law fluid, we achieve a finite viscosity as the shear rate gets zero $(\dot{\gamma} = 0)$ and secondly it involves a time constant due to which it is good enough for many engineering calculations. Examples of the applications of the Cross model includes the synthesis of the polymeric solutions like 0.35% aqueous solution of Xanthan gum, blood, aqueous solution of polymer latex sphere, 0.4% aqueous solution of polyacrylamide [7]. The Cross rheology equation [8-9] for viscosity in terms of shear rate is given as:

$$\eta^* = \eta_\infty + (\eta_0 - \eta_\infty)\left[\frac{1}{1+(\Gamma\dot{\gamma})^{1-n}}\right], \tag{1}$$

or equivalently,

$$\frac{\eta_0 - \eta^*}{\eta^* - \eta_\infty} = (\Gamma\dot{\gamma})^{1-n}, \tag{2}$$

where $\eta_0$ and $\eta_\infty$ are the limiting viscosities at low and high shear rates, respectively. $\Gamma$ is the material time constant known as the consistency index, $n$ the dimensionless constant, commonly known as the flow behavior index and $\dot{\gamma}\left(=\frac{1}{2}\Pi\right)$ the shear rate with $\Pi$ the second invariant strain rate tensor. Cross [6] presented an experimental data for many systems by using a simple value $n = 2/3$ but he clearly stated that there is no hindrance in treating $n$ as an adjustable parameter [10]. This model predicts the usual Newtonian fluid if $n=1$ or $\Gamma=0$ or both.

---
[1]Corresponding author. E-mail address: mehwish.manzur@gmail.com

It is interesting to note that by making certain approximation to the Cross equation, we can achieve various other popular viscosity models like the power-law model, the Sisko model and the Bingham model. When $\eta^* \ll \eta_0$ and $\eta^* \gg \eta_\infty$, the Cross equation (2) reduces to

$$\eta^* = K_1(\dot{\gamma})^{n-1}, \tag{3}$$

which is the well-known power-law model with $K_1$ the consistency index and $n$ the power-law index.

Furthermore, if $\eta^* \ll \eta_0$, we get

$$\eta^* = \eta_\infty + K_1(\dot{\gamma})^{n-1}, \tag{4}$$

which is the renowned Sisko rheological model.
Further, by setting $n=1$ in Sisko model and by slight redefinition of parameters we attain

$$\eta^* = \mu_B + \sigma_0\dot{\gamma}, \tag{5}$$

which represents the Bingham model [11] such that $\mu_B$ is the Bingham plastic viscosity and $\sigma_0$ is the Bingham yield stress.

In the past two decades experimental based study is done on the Cross model by several investigators. Escudier *et al*. [12] performed the experimental analysis and presented the fluid-flow data by fitting the Cross model to the non-Newtonian liquid particularly the Xanthan gum (XG). Xie and Jin [13] investigated the Cross rheology equation to analyze the free surface flow of non-Newtonian fluids. For the numerical implementation of the Cross equation an experimental based method namely the WC-MPS method was employed to determine the four rheology parameters of the Cross model.

The analysis of the flow and heat transfer over a continuously stretching surface has acquired great importance due to its occurring in many engineering and industrial applications like drawing of rubber and plastic sheets, polymer processing and metallurgy, crystal growing, food processing and many others. The rate of cooling and stretching renders significant part in controlling the quality of the final product. The boundary layer flow over a moving surface with constant speed was initiated by Sakiadis [14]. Later, Crane [15] extended this work for the case of linearly stretching sheet. After these pioneering attempts, research on the flow and heat transfer over a stretching surface has immensely been done by various investigators [16-19].

A thorough survey of the literature reveals that the Cross fluid which predicts the pseudoplastic and dilatant nature of the fluid over a wide range of shear rate has not been given due attention. In fact to the best of authors knowledge no attempt has been made on the boundary layer flow and heat transfer of the Cross fluid. The present work fills this gap by presenting the boundary layer equations for the flow of Cross fluid over a linearly stretching sheet. The governing equations are numerically solved by the help of the shooting technique and effect of the pertinent parameters is revealed through graphs and tables.

## 2. Governing equations

The conservation equations of mass, linear momentum and energy for the flow of an incompressible fluid are

$$\text{div } \mathbf{V} = 0, \tag{6}$$

$$\rho \frac{d\mathbf{V}}{dt} = \text{div } \boldsymbol{\tau}, \tag{7}$$

$$\rho c_p \frac{dT}{dt} = \boldsymbol{\tau}.\mathbf{L} - \text{div } \mathbf{q}, \tag{8}$$

where $\mathbf{V}$ symbolizes the velocity vector, $\rho$ the density, $\boldsymbol{\tau}$ the Cauchy stress tensor, $T$ the fluid temperature, $c_p$ the specific heat at constant pressure, $\mathbf{q}\,(=-k\nabla T)$ the heat flux vector, $\mathbf{L} = \nabla \mathbf{V}$ and $\frac{d}{dt}$ the material derivative.

The Cauchy stress tensor for the four parameter fluid is defined as

$$\boldsymbol{\tau} = -p\mathbf{I} + \eta^* \mathbf{A}_1, \tag{9}$$

where $\eta^*$ for the Cross fluid is given by Eq. (1), where

$$\mathbf{A}_1 = \mathbf{L} + \mathbf{L}^T, \quad \Pi = \sqrt{\frac{1}{2} tr(\mathbf{A}_1^2)}, \tag{10}$$

such that $p$ is the pressure, $\mathbf{I}$ denotes the identity tensor and $\mathbf{A}_1$ the first Rivlin-Ericksen tensor.

The infinite shear rate viscosity $\eta_\infty$ is frequently set equal to zero [20-22] in Eq. (1) and accordingly Eq. (9) reduces to

$$\boldsymbol{\tau} = -p\mathbf{I} + \eta_0 \left[ \frac{1}{1+(\Gamma \dot{\gamma})^{1-n}} \right] \mathbf{A}_1. \tag{11}$$

The salient feature of the Cross model is that when $0 < n < 1$ the fluid is shear-thickening while if $n > 1$, the fluid is shear-thinning. Additionally, when $n = 1$, it reduces to the usual Newtonian fluid.

For a two-dimensional flow in Cartesian coordinates, we assume the velocity and temperature fields of the form

$$\mathbf{V} = [u(x, y), v(x, y), 0], \quad T = T(x, y), \tag{12}$$

where $u$ and $v$ are the $x-$ and $y-$components of the velocity vector.

Keeping in view Eq. (12), the shear rate $\dot{\gamma}$ is expressed as:

$$\dot{\gamma} = \left[ 4\left(\frac{\partial u}{\partial x}\right)^2 + \left(\frac{\partial u}{\partial y} + \frac{\partial v}{\partial x}\right)^2 \right]^{\frac{1}{2}}. \tag{13}$$

Substituting Eq. (12) in Eqs. (6) and (7), bearing in mind Eqs. (11) and (13), a straight forward calculation yields the following governing equations

$$\frac{\partial u}{\partial x} + \frac{\partial v}{\partial y} = 0, \tag{14}$$

$$\rho\left(u\frac{\partial u}{\partial x} + v\frac{\partial u}{\partial y}\right) = -\frac{\partial p}{\partial x} + 2\eta_0 \frac{\partial}{\partial x}\left[\frac{\frac{\partial u}{\partial x}}{1 + \left\{\Gamma^2\left(4\left(\frac{\partial u}{\partial x}\right)^2 + \left(\frac{\partial u}{\partial y} + \frac{\partial v}{\partial x}\right)^2\right)\right\}^{\frac{1-n}{2}}}\right]$$

$$+ \eta_0 \frac{\partial}{\partial y}\left[\frac{\left(\frac{\partial u}{\partial y} + \frac{\partial v}{\partial x}\right)}{1 + \left\{\Gamma^2\left(4\left(\frac{\partial u}{\partial x}\right)^2 + \left(\frac{\partial u}{\partial y} + \frac{\partial v}{\partial x}\right)^2\right)\right\}^{\frac{1-n}{2}}}\right], \tag{15}$$

$$\rho\left(u\frac{\partial v}{\partial x} + v\frac{\partial v}{\partial y}\right) = -\frac{\partial p}{\partial y} + \eta_0 \frac{\partial}{\partial x}\left[\frac{\left(\frac{\partial u}{\partial y} + \frac{\partial v}{\partial x}\right)}{1 + \left\{\Gamma^2\left(4\left(\frac{\partial u}{\partial x}\right)^2 + \left(\frac{\partial u}{\partial y} + \frac{\partial v}{\partial x}\right)^2\right)\right\}^{\frac{1-n}{2}}}\right]$$

$$+ 2\eta_0 \frac{\partial}{\partial y}\left[\frac{\frac{\partial v}{\partial y}}{1 + \left\{\Gamma^2\left(4\left(\frac{\partial u}{\partial x}\right)^2 + \left(\frac{\partial u}{\partial y} + \frac{\partial v}{\partial x}\right)^2\right)\right\}^{\frac{1-n}{2}}}\right], \tag{16}$$

The above equations of motion are made non-dimensional through the following relations

$$(x, y) = L(x^*, y^*), (u, v) = U(u^*, v^*) \text{ and } p = \rho U^2 p^*. \tag{17}$$

In terms of the dimensionless variables, the continuity and momentum equations take the following forms:

$$\frac{\partial u^*}{\partial x^*} + \frac{\partial v^*}{\partial y^*} = 0, \tag{18}$$

$$u^* \frac{\partial u^*}{\partial x^*} + v^* \frac{\partial u^*}{\partial y^*} = -\frac{\partial p^*}{\partial x^*} + 2\varepsilon_1 \frac{\partial}{\partial x^*}\left[\frac{\frac{\partial u^*}{\partial x^*}}{1+\left\{\varepsilon_2\left(4\left(\frac{\partial u^*}{\partial x^*}\right)^2 + \left(\frac{\partial u^*}{\partial y^*}+\frac{\partial v^*}{\partial x^*}\right)^2\right)\right\}^{\frac{1-n}{2}}}\right]$$

$$+\varepsilon_1 \frac{\partial}{\partial y^*}\left[\frac{\left(\frac{\partial u^*}{\partial y^*}+\frac{\partial v^*}{\partial x^*}\right)}{1+\left\{\varepsilon_2\left(4\left(\frac{\partial u^*}{\partial x^*}\right)^2 + \left(\frac{\partial u^*}{\partial y^*}+\frac{\partial v^*}{\partial x^*}\right)^2\right)\right\}^{\frac{1-n}{2}}}\right], \quad (19)$$

$$u^* \frac{\partial v^*}{\partial x^*} + v^* \frac{\partial v^*}{\partial y^*} = -\frac{\partial p^*}{\partial y^*} + \varepsilon_1 \frac{\partial}{\partial x^*}\left[\frac{\left(\frac{\partial u^*}{\partial y^*}+\frac{\partial v^*}{\partial x^*}\right)}{1+\left\{\varepsilon_2\left(4\left(\frac{\partial u^*}{\partial x^*}\right)^2 + \left(\frac{\partial u^*}{\partial y^*}+\frac{\partial v^*}{\partial x^*}\right)^2\right)\right\}^{\frac{1-n}{2}}}\right]$$

$$+2\varepsilon_1 \frac{\partial}{\partial y^*}\left[\frac{\frac{\partial v^*}{\partial y^*}}{1+\left\{\varepsilon_2\left(4\left(\frac{\partial u^*}{\partial x^*}\right)^2 + \left(\frac{\partial u^*}{\partial y^*}+\frac{\partial v^*}{\partial x^*}\right)^2\right)\right\}^{\frac{1-n}{2}}}\right], \quad (20)$$

where the dimensionless parameters are defined as:

$$\varepsilon_1 = \frac{\frac{\eta_0}{\rho}}{LU} \text{ and } \varepsilon_2 = \frac{\Gamma^2}{\left(\frac{L}{U}\right)^2}. \quad (21)$$

Using the standard boundary layer assumptions, where $x$, $u$, $p$ are of order 1 while $v$ and $y$ are of order $\delta$. The dimensionless parameters $\varepsilon_1$ and $\varepsilon_2$ are of order $\delta^2$.

Keeping in view the boundary layer analysis, one finally has the following equations in dimensional form

$$\frac{\partial u}{\partial x} + \frac{\partial v}{\partial y} = 0, \quad (22)$$

$$u \frac{\partial u}{\partial x} + v \frac{\partial u}{\partial y} = -\frac{1}{\rho}\frac{\partial p}{\partial x} + \nu \frac{\partial}{\partial y}\left[\frac{\frac{\partial u}{\partial y}}{1+\left\{\Gamma\left(\frac{\partial u}{\partial y}\right)\right\}^{1-n}}\right], \quad (23)$$

where $\nu = \frac{\eta_0}{\rho}$ gives the kinematic viscosity.

## 3. Problem formulation

The steady two-dimensional flow and heat transfer of an incompressible generalized Newtonian fluid specifically the Cross fluid is considered by incorporating the effects of stretching surface. The stretching sheet is taken to be coinciding with the plane $y = 0$ while the fluid occupies the region $y > 0$. The sheet is uniformly stretched along the $x$-axis with a velocity $U = cx$, where $c\ (>0)$ is the stretching rate of the sheet. We assume that the temperature near the surface of the sheet is $T_w$ while $T_\infty$ is the ambient fluid temperature. Moreover, for the present case the driving pressure gradient is not present and the flow is driven only due to stretching sheet.

Under these assumptions, the system of equations governing the flow and heat transfer is

$$\frac{\partial u}{\partial x} + \frac{\partial v}{\partial y} = 0, \tag{24}$$

$$u\frac{\partial u}{\partial x} + v\frac{\partial u}{\partial y} = \upsilon \frac{\partial}{\partial y}\left[\frac{\frac{\partial u}{\partial y}}{1+\{\Gamma(\frac{\partial u}{\partial y})\}^{1-n}}\right], \tag{25}$$

$$u\frac{\partial T}{\partial x} + v\frac{\partial T}{\partial y} = \alpha \frac{\partial^2 T}{\partial y^2}, \tag{26}$$

where $\alpha \left(= \frac{k}{\rho c_p}\right)$ is the thermal diffusivity and $k$ being the thermal conductivity and $c_p$ the specific heat at constant pressure.

The relevant boundary conditions of the problem under consideration are:

$$u(x, y) = U(x) = cx, \ v(x, y) = 0, \ T(x, y) = T_w \quad \text{at } y = 0, \tag{27}$$

$$u(x, y) \to 0, \ T(x, y) \to T_\infty \quad \text{as } y \to \infty. \tag{28}$$

Employing the following local similarity transformations

$$\eta = \sqrt{\frac{c}{v}} y, \ \psi = \sqrt{vc}\, x f(\eta), \ \theta(\eta) = \frac{T - T_\infty}{T_w - T_\infty}, \tag{29}$$

where $\eta$ denotes the dimensionless local similarity variable and $\psi$ the stream function such that $(u, v) = \left(\frac{\partial \psi}{\partial y}, -\frac{\partial \psi}{\partial x}\right)$. In consideration of the above transformations, the incompressibility condition (24) is automatically satisfied while Eqs. (25-28) are reduced to

$$\left[ff'' - (f')^2\right]\left[1 + (We f'')^{1-n}\right]^2 + \left[1 + n(We f'')^{1-n}\right]f''' = 0, \tag{30}$$

$$\theta'' + \Pr f\theta' = 0, \tag{31}$$

$$f(\eta) = 0, \ f'(\eta) = 1, \ \theta(\eta) = 1 \quad \text{at } \eta = 0, \tag{32}$$

$$f'(\eta) \to 0, \ \theta(\eta) \to 0 \quad \text{as } \eta \to \infty, \tag{33}$$

where prime denotes differentiation with respect to $\eta$, $We$ stands for the local Weissenberg number and $\Pr$ the Prandtl number defined as:

$$We = c\Gamma \operatorname{Re}^{\frac{1}{2}}, \ \Pr = \frac{\eta_0 c_p}{k}, \tag{34}$$

such that $\operatorname{Re}\left(= \frac{cx^2}{v}\right)$ gives the local Reynolds number.

The boundary layer equations for the flow and heat transfer of Newtonian fluid can be achieved by setting $n = 1$ or $We = 0$ in Eqs. (30-33).

The expression for the local skin friction coefficient and the local Nusselt number is given by

$$C_f = \frac{\tau_w}{\frac{1}{2}\rho U^2}, \ N_u = \frac{xq_w}{k(T_w - T_\infty)}, \tag{35}$$

where $\tau_w$ signifies the local wall shear stress and $q_w$ the surface heat flux defined as:

$$\tau_w = \tau_{xy}\big|_{y=0} = \left[\eta_0 \frac{\frac{\partial u}{\partial y}}{1+\left\{\Gamma\left(\frac{\partial u}{\partial y}\right)\right\}^{1-n}}\right]_{y=0}, \quad q_w = -k\frac{\partial T}{\partial y}\bigg|_{y=0}. \tag{36}$$

In view of Eq. (29), the dimensionless forms of the local skin friction coefficient and the local Nusselt number can be obtained as

$$\frac{1}{2}\text{Re}^{\frac{1}{2}} C_{f_x} = \frac{f''(0)}{1+\left(We f''(0)\right)^{1-n}}, \quad -\text{Re}^{-\frac{1}{2}} Nu_x = \theta'(0). \tag{37}$$

## 4. Results and discussion

The non-linear differential equations (30) and (31) subject to boundary conditions (32) and (34) are numerically solved by the help of shooting method. The influence of the emerging parameters is graphically examined and discussed in this section. The authenticity of the obtained numerical results is verified by making comparison with the existing literature. The numerical values of the local skin friction coefficient and the local Nusselt number are also tabulated.

The impact of the power-law index $n$ on the velocity and temperature fields is respectively shown in figures 1(a) and 1(b) for the case of shear-thickening $(n<1)$ as well as shear-thinning fluid $(n>1)$. An analysis of figure 1(a) shows that the incremented values of $n$ diminishes the fluid velocity for shear-thickening fluid while an enlargement in the velocity is seen for shear-thinning fluid. Figure 1(b) discloses that a boost in the temperature field is seen for shear-thickening regime while a declining trend is observed for shear-thinning regime. Furthermore, the momentum and thermal boundary layers thickness is larger for shear-thickening fluid in comparison with shear-thinning fluid. The physical reason behind this trend is that shear-thickening fluids bestow more resistance to flow because of high viscosity. Hence the velocity of the fluid decreases while the fluid temperature rises. Opposite trend is visualized for shear-thinning fluid which renders less resistance to fluid motion due to less viscosity and due to which the velocity profile increases and the fluid temperature decreases.

Figures 2 and 3 are sketched to demonstrate the behavior of the velocity and temperature profiles corresponding to a change in local Weissenberg number $We$. Also comparative plots are provided for shear- thickening $(n<1)$ as well as shear-thinning $(n>1)$ profiles. Figure 2(a-b) exhibits that the velocity profiles as well as the momentum boundary layer thickness are lowered by increasing the Weissenberg number. An elevation in the temperature and the thermal boundary layer is noticed for growing values of $We$ as displayed in figure 3(a-b). Moreover, a careful examination of these figures discloses that this behavior is much prominent for shear-thickening fluids $(n=0.5)$. From the physical point of view, an increase in $We$ causes an enlargement in the relaxation time which results in lowering the fluid velocity and escalating the fluid temperature.

Figure 4 is plotted to see the dependence of the temperature profile on the Prandtl number $Pr$ for shear-thinning $(n=1.5)$ as well as shear-thickening $(n=0.5)$ regimes. The growing values of the Prandtl number results in the reduction of the temperature distribution and thermal boundary layer thickness. Physically, the Prandtl number demonstrates the ratio of the momentum diffusivity to thermal diffusivity. For elevated values of $Pr$, the thermal diffusivity gets weaker as a result the flow of heat into the fluid is restrained and thermal boundary layer structures gets diminished. Heat diffuses faster from the wall for the fluids having low Prandtl number due to high thermal conductivity. Thus, Prandtl number acts as a controlling factor in conducting flows for monitoring the rate of cooling.

In order to authenticate the present results a comparative study is made with the existing results in the literature. Table 1 is constructed to provide comparison of the value of $-f''(0)$ with the results of Cortell [21], Cortell [22] and Hamad and Ferdows [23] for the case of viscous fluid. A comparison with the work of Wang [24], Gorla and Sidawi [25] and Hamad [26] is given in Table 2. The numerical values of the local Nusselt number is calculated for the limiting case $(We=0)$. An excellent compatibility with the existing literature is thus acheived.

The numerical values of the local skin friction coefficient and the local Nusselt number are tabulated in tables 3 and 4, respectively. The consequence of the power-law index $n$ and the local Weissenberg number $We$ on the local skin friction coefficient is displayed in Table 3. It is inferred from the data that the magnitude of the skin friction coefficient is a decreasing function of the Weissenberg number for fixed value of $n$. Further, it is seen that the magnitude of the local skin friction coefficient is raised for shear-thickening fluid $(n<1)$ while opposite trend is observed for shear-thinning fluid $(n>1)$. The influence of the power-law index $n$, local Weissenberg number $We$ and the Prandtl number $Pr$ on the local Nusselt number is provided in Table 4 by presenting numerical data. It is deduced that the local Nusselt number enlarges with augmenting values of the Prandtl number $Pr$. This is due to the fact that the growing values of $Pr$ enhances the process of convection in comparison with the conduction due to which the rate of heat transfer increases. Opposite trend is revealed for the case of growing values of the local Weissenberg number $We$. It is also noticed that the rate of heat transfer lessens with a rise in the values of the power-law index $n$ for the case of shear- thickening fluid $(n<1)$. However, the rate of heat transfer is enhanced for shear-thinning fluid $(n>1)$.

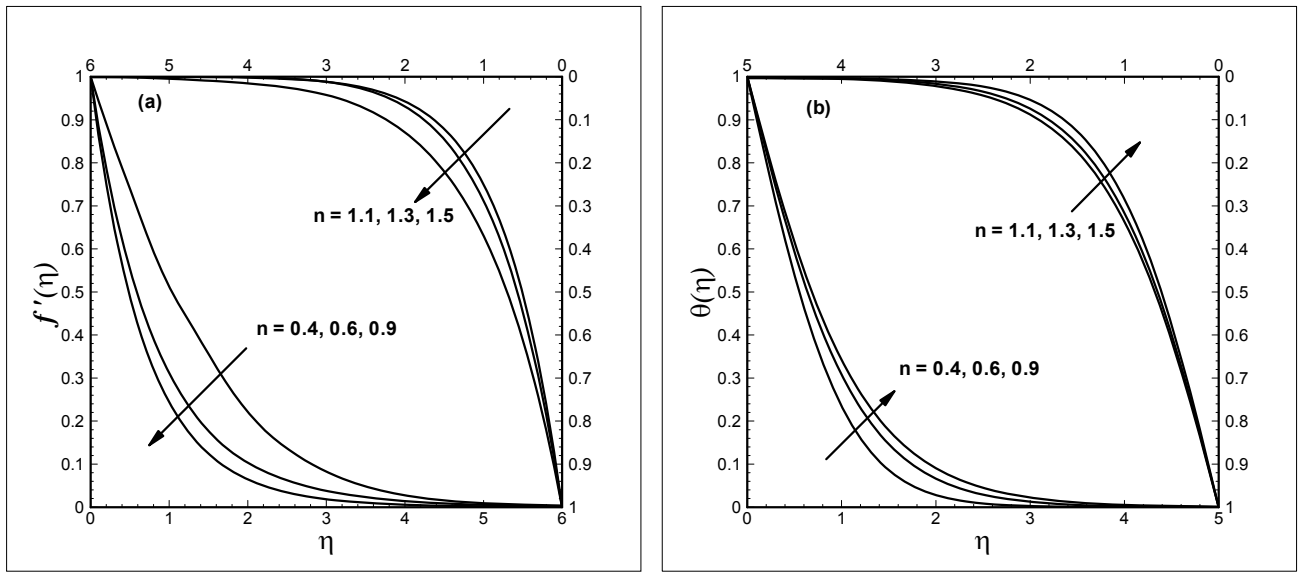

**Figure 1: Velocity and temperature profiles for different values of the power law index $n$ when $We=2$ and $Pr=5$.**

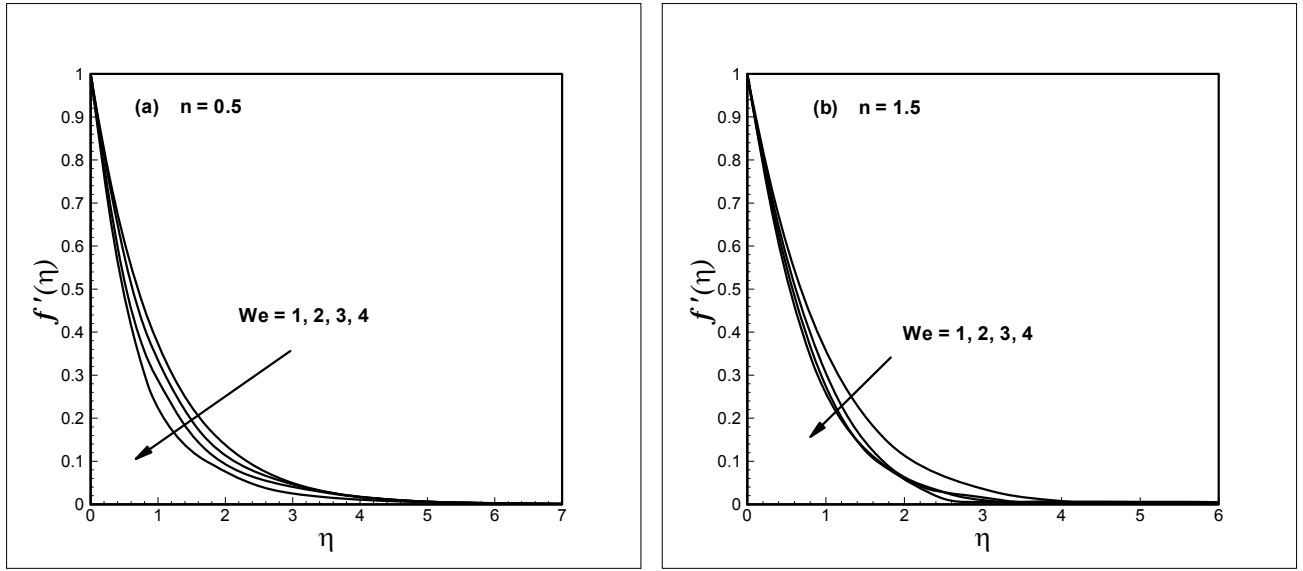

**Figure 2: Velocity and temperature profiles for different values of the local Weissenberg number** $We$**.**

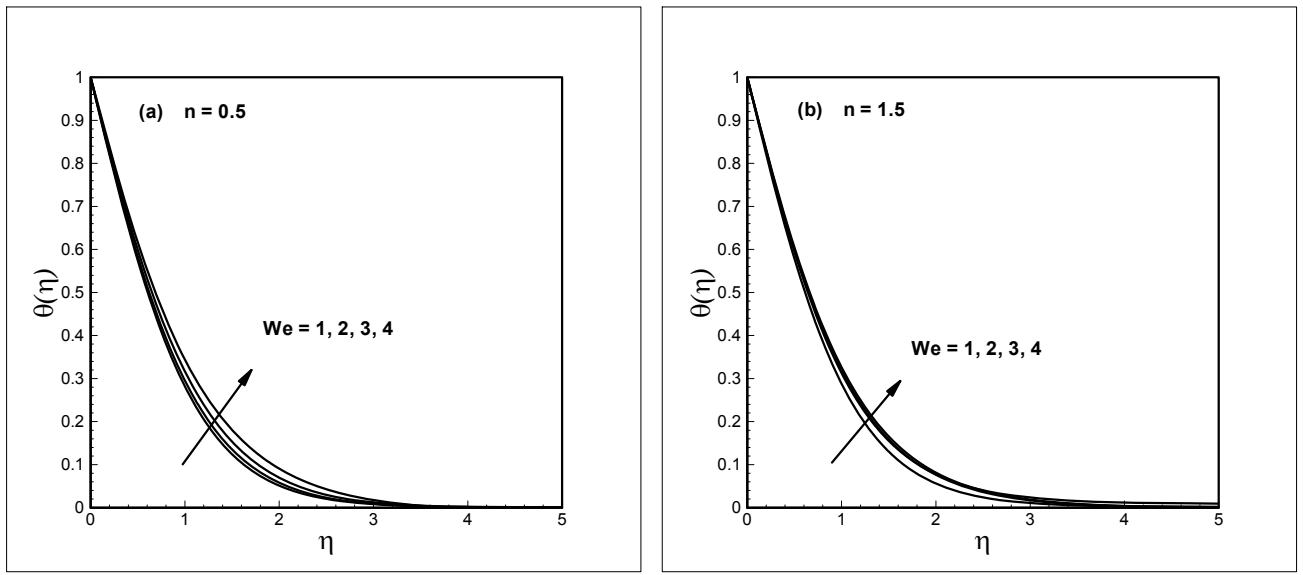

**Figure 3: Velocity and temperature profiles for different values of the local Weissenberg number** $We$ **when** $\Pr = 5$.

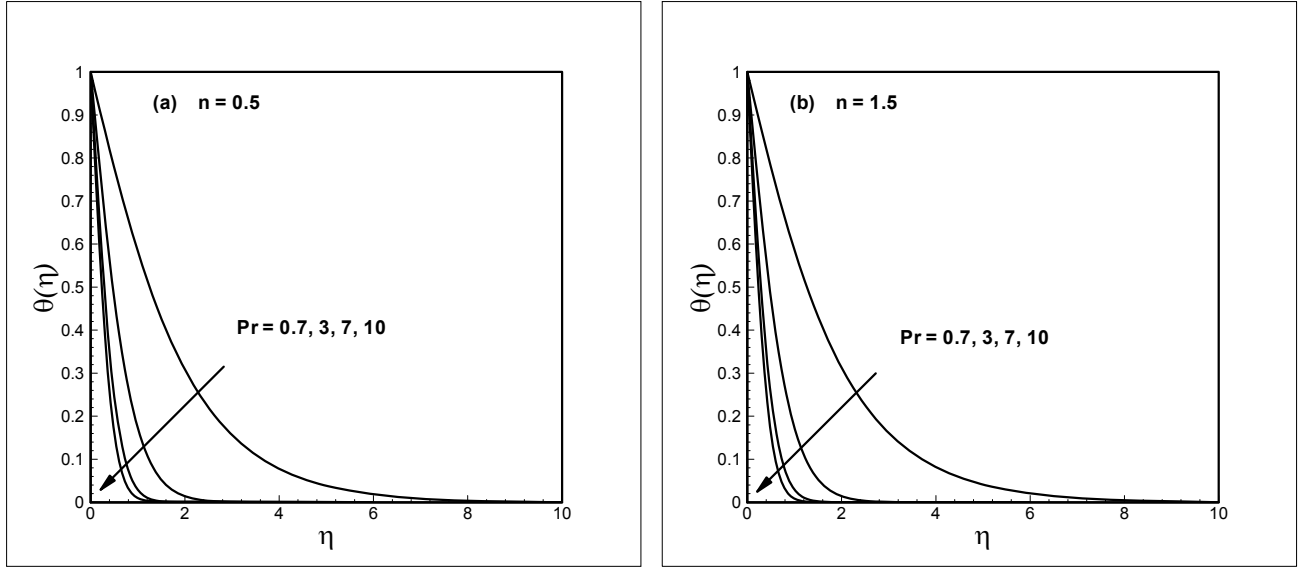

**Figure 4:** Velocity and temperature profiles for different values of the Prandtl number Pr when $We = 2$.

**Table 1: Comparison of the values of $-f''(0)$ for the case of the Newtonian fluid $(We = 0)$.**

|  | Cortell [21] | Cortell [22] | Hamad and Ferdows [23] | Present result |
|---|---|---|---|---|
| $-f''(0)$ | 1.0 | 1.0 | 1.0043 | 1.00001 |

**Table 2: Comparison of the variation of $-\theta'(0)$ for the Newtonian fluid $(We = 0)$.**

| Pr | $-\theta'(0)$ | | | |
|---|---|---|---|---|
|  | Wang [24] | Gorla and Sidawi [25] | Hamad [26] | Present results $(We = 0)$ |
| 0.07 | 0.0656 | 0.0656 | 0.0656 | 0.065526 |
| 0.2 | 0.1691 | 0.1691 | 0.1691 | 0.164037 |
| 0.7 | 0.4539 | 0.4539 | 0.4539 | 0.418299 |
| 2 | 0.9114 | 0.9114 | 0.9114 | 0.826827 |
| 7 | 1.8954 | 1.8905 | 1.8954 | 1.80433 |
| 20 | 3.3539 | 3.3539 | 3.3539 | 3.25603 |
| 70 | 6.4622 | 6.4622 | 6.4622 | 6.36662 |

**Table 3**: Numerical values of the local skin friction coefficient $\left(-\frac{1}{2}\text{Re}^{\frac{1}{2}} C_f\right)$ for various values of $We$ and $n$.

| Parameters (fixed values) | Parameters | | $-\frac{1}{2}\text{Re}^{\frac{1}{2}} C_f$ |
|---|---|---|---|
| $We = 2$ | $n$ | 0.4 | 0.226604 |
| | | 0.6 | 0.443713 |
| | | 0.9 | 0.684548 |
| | | 1.1 | 0.712246 |
| | | 1.3 | 0.698608 |
| | | 1.5 | 0.666879 |
| $n = 0.5$ | $We$ | 1 | 0.483845 |
| | | 2 | 0.322239 |
| | | 3 | 0.241925 |
| | | 4 | 0.184615 |

**Table 4**: Numerical values of the local Nusselt number $\left(-\text{Re}^{\frac{1}{2}} Nu_x\right)$ for various values of $We$, $n$ and $\text{Pr}$.

| Parameters (fixed values) | Parameters | | $-\text{Re}^{\frac{1}{2}} Nu_x$ |
|---|---|---|---|
| $We = 2$, $\text{Pr} = 5$ | $n$ | 0.4 | 0.991493 |
| | | 0.6 | 0.868297 |
| | | 0.9 | 0.821454 |
| | | 1.1 | 0.830717 |
| | | 1.3 | 0.863791 |
| | | 1.5 | 0.910371 |
| $n = 0.5$, $\text{Pr} = 5$ | $We$ | 1 | 0.915391 |
| | | 2 | 0.914499 |
| | | 3 | 0.906314 |
| | | 4 | 0.897812 |
| $n = 0.5$, $We = 2$ | $\text{Pr}$ | 0.7 | 0.46173 |
| | | 3 | 1.16766 |
| | | 7 | 1.90855 |
| | | 10 | 2.31302 |

# 5. Concluding remarks

The present study is an inaugural work in bestowing the boundary layer equations for the flow of the Cross fluid. The modelled partial differential equations were converted to ordinary differential equations by the help of suitable local similarity transformations and then numerically integrated by employing the shooting technique. The graphs were constructed for the velocity and

temperature fields corresponding to the emerging parameters. The main findings are summarized as:

- The uplifting values of the power-law index in the shear- thickening regime $(n<1)$, lowered the momentum boundary layer thickness while a boost in the thermal boundary layer was observed. However, reverse behavior was obtained for the case of shear thinning fluid $(n>1)$.
- A decline in the velocity profile was visualized for growing values of the local Weissenberg number. However, the temperature profiles exhibited a progressive trend with much prominent effects for the case of shear-thickening fluid $(n<1)$.
- The escalated values of the Prandtl number lessened the temperature profile as well as the thermal boundary layer thickness.
- It was seen that the magnitude of the local skin friction coefficient decreased for growing values of the local Weissenberg number where as the local Nusselt number boosted with increase in $We$.